\documentclass[
 floatfix,
 reprint,
superscriptaddress,
 amsmath,amssymb,
 aps,
prb,
]{revtex4-2}

\usepackage{graphicx}
\usepackage{dcolumn}
\usepackage{bm}
\usepackage{multirow}
\usepackage{float}

\def\KMAO{K$_2$Mn$_3$(AsO$_4$)$_3$}
\def\ECA{EuCd$_2$As$_2$}

\def\AFMPO{Ag$_2$FeMn$_2$(PO$_4$)$_3$}
\def\NCFPO{Na$_4$CaFe$_4$(PO$_4$)$_6$}

\def\degrees{$^{\circ}$}
\def\iA{$\text{\AA}^{-1}$}

\def\Ctc{$C2/c$}

\def\k{$\mathbf{k}$}
\def\ko{$\mathbf{k_1}$}
\def\kt{$\mathbf{k_2}$}

\begin{document}

\preprint{APS/123-QED}

\title{Insulating ground state and 2-k magnetic structure of candidate Weyl Hydrogen atom K$_2$Mn$_3$(AsO$_4$)$_3$}

\author{Keith M. Taddei}
 \email[corresponding author ]{ktaddei@anl.gov}
 \affiliation{X-ray Sciences Division, Advanced Photon Source, Argonne National Laboratory, Lemont, Illinois 60439, USA}
 \affiliation{Neutron Scattering Division, Oak Ridge National Laboratory, Oak Ridge, Tennessee 37834, USA}
\author{Kulugammana G. S. Ranmohotti}%
\affiliation{Governors State University, 1 University Pkwy, University Park, Illinois 60484, USA}%
\author{Duminda S. Liurukara}
\affiliation{
 Neutron Sciences Directorate, Oak Ridge National Laboratory, Oak Ridge, Tennessee 37832, USA}
 \author{Alex Martinson}
\affiliation{
 Materials Science Division, Argonne National Laboratory, Lemont, Illinois 60439, USA}
\author{Stuart Calder}
\affiliation{Neutron Scattering Division, Oak Ridge National Laboratory, Oak Ridge, Tennessee 37834, USA}
\author{German Samolyuk}
\affiliation{
 Materials Science and Technology Division, Oak Ridge National Laboratory, Oak Ridge, Tennessee 37834, USA}
 \author{Nabaraj Pokhrel}
\affiliation{
 Materials Science and Technology Division, Oak Ridge National Laboratory, Oak Ridge, Tennessee 37834, USA}
\author{Daniel Phelan}
\affiliation{
 Materials Science Division, Argonne National Laboratory, Lemont, Illinois 60439, USA}
\author{David Parker}
\affiliation{
 Materials Science and Technology Division, Oak Ridge National Laboratory, Oak Ridge, Tennessee 37834, USA}

\date{\today}

\begin{abstract}
The ideal Weyl \lq Hydrogen-atom\rq\ semi-metal exhibits only a single pair of Weyl nodes and no other trivial states at the Fermi energy. Such a material would be a panacea in the study of Weyl quasi particles allowing direct unambiguous observation of their topological properties. The alluaudite-like \KMAO\ compound was recently proposed as such a material. Here we use comprehensive experimental work and first principle calculations to assess this prediction. We find \KMAO\ crystallizes in the \Ctc\ symmetry with a quasi-1D Mn sublattice, growing as small needle-like crystals. Bulk properties measurements reveal magnetic transitions at $\approx$ 8 and $\approx$ 4 K which neutron scattering experiments show correspond to two distinct magnetic orders, first a partially ordered ferrimagnetic \ko$=(0,0,0)$ structure at 8 K and a second transition of \kt$=(1,0,0)$ at 4 K to a fully ordered state. Below the second transition, both ordering vectors are necessary to describe the complex magnetic structure with modulated spin magnitudes. Both of the best-fit magnetic structures in this work are found to break the symmetry necessary for the generation of the Weyl nodes, though one of the magnetic structures allowed by \ko\ does preserve this symmetry. However, the crystals are optically transparent and ellipsometry measurements reveal a large band-gap, undermining expectations of semi-metallic behavior. Density functional theory calculations predict an insulating antiferromagnetic ground state, in contrast to previous reports, and suggest potential frustration on the magnetic sublattice. Given the wide tunability of the alluaudite structure we consider ways to push the system closer to semi-metallic state. 

\end{abstract}

\maketitle


\section{\label{sec:Intro}Introduction}

Weyl semi-metals offer experimental access to a massless chiral solution of the relativistic wave equation which has yet to be observed in the standard model and in doing so, manifest topological properties useful for new paradigms of device design \cite{Dirac1928a, Herring1937, Hasan2017}. These materials, host a band inversion defined by a linearly dispersing four-fold degenerate band crossing (a Dirac point) whose degeneracy is subsequently broken via a splitting of its spin-component \cite{Vafek2014}. This generates a series of Weyl points; chiral nodes which act as mono-polar sources and sinks of Berry curvature. These Weyl points and their associated quasi-particles are robust - they can only be gapped by bringing two opposite chirality Weyl points to the same position in the Brillouin zone - and generate novel bulk properties such as the chiral anomaly, unclosed Fermi surface arcs, non-saturating magnetization, and even exhibit Axion electrodynamics \cite{Armitage2018, Xu2015, Zhang2019,Zyuzin2012} . This combination of properties and topological protection enables potentially powerful new designs across a wide breadth of technologies - from catalysis to quantum computing - and has thus generated significant efforts towards experimentally realizing and optimizing Weyl semi-metals \cite{Osterhoudt2019,Ukhtary2022,Chen2021,He2021}.         

Though the prediction of condensed matter Weyl quasi-particles came shortly after the prediction of Weyl fermions, only recently have they been successfully found experimentally in materials. This was first achieved with the prediction and verification of TaAs as a Weyl semi-metal which hosts several signatures of Weyl quasiparticles, such as Fermi arcs, chiral Fermi pockets, topological surface states, and the chiral anomaly \cite{Lv2015, Belopolski2016, Arnold2016, Batabyal2016, Huang2015}. Since this initial discovery, numerous additional Weyl materials have been discovered and new properties engendered of the Weyl quasi-particles have arisen such as the planar Hall effect in GdPtBi, giant magneto-optical response in Co$_3$Sn$_2$S$_2$, higher order Weyl phases in (TaSe$_4$)$_2$I, helical photocurrents in RhSi and the anomalous Nernst effect in \ECA \cite{Kumar2018, Okamura2020, Wei2024, Rees2020, Li2021, Xu2021}. 

However, considerable challenges remain in the study of Weyl quasi-particles due to the complex band structures that most known Weyl semi-metals exhibit, which often host either numerous pairs of Weyl nodes or additional trivial band crossings - or both. This situation is partially due to the preponderance of Weyl materials whose Weyl points are borne of inversion symmetry breaking (ISB). Such materials are forced, by symmetry, to have no fewer than four Weyl nodes (two opposite chirality sets) which are not constrained to exist at the same energy in the band structure. For example in the ISB TaAs, this leads to the formation of twelve pairs of Weyl nodes some of which straddle the Fermi energy \cite{Lv2015, Xu2015}. Such a complicated band structure begets higher order interactions between the Weyl quasi-particles and invariably leads to trivial band crossings which add non-Weyl carriers that dilute the topological transport properties. 

To avoid this situation, one can instead find materials in which time-reversal symmetry breaking (TRSB) splits the spin degeneracy. Such a scenario leads to the minimum possible number of Weyl nodes allowed by the Nielsen-Ninomiya \lq no-go\rq\ theorem - two, one of each chirality - which are constrained by the remaining IS to be at the same energy \cite{Nielsen1981I,Nielsen1981II}. This allows the possibility of a band structure with only two Weyl nodes both exactly at the Fermi level and no additional trivial band crossings. Such a configuration has been dubbed the Weyl \lq Hydrogen atom\rq\ and is the simplest realization of Weyl physics in the condensed matter setting. A Weyl Hydrogen atom material will give rise to the clearest signatures of Weyl physics with all transport deriving exclusively from the Weyl quasi-particles and so is highly desirable for both testing and harnessing their exotic properties. However, thus far, relatively few candidate Weyl Hydrogen atom materials have been proposed, and even fewer experimentally studied, due to the difficulty in finding such a band structure in a material which also realizes a magnetic order that does not gap out the band crossing \cite{Liu2024, Xu2011, Gao2023}. 

\begin{figure}
\includegraphics[width=\columnwidth]{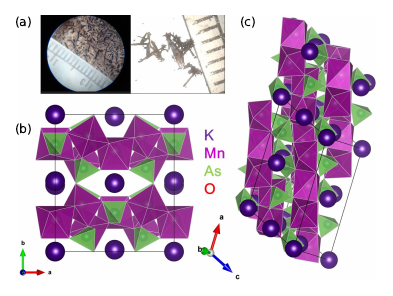}
\caption{\label{fig:struct} (a) Images of as grown single crystals shown with mm scale for reference. Crystal structure of \KMAO\ shown (b) along the \textit{\textbf{c}} axis and (c) oriented to exhibit the Mn chain sublattice. K, Mn, and As atoms are shown in purple, magenta, and green. The O atoms are not shown to help with visual clarity but are positioned at the vertices of the polyhedron.}
\end{figure}

Recently, \ECA\ has received both theoretical and experimental support as a potential Weyl Hydrogen atom. Band structure calculations predicted the ideal Weyl configuration for a simple ferromagnetic structure and numerous reports have shown experimental evidence of topological transport \cite{Ma2019, Rahn2018,Soh2019,Wang2019}. Furthermore, its experimentally determined ground state magnetic structure was shown to generate a pair of Weyl nodes and trigger the chiral anomaly in transport measurements strongly suggesting a Weyl state \cite{Taddei2022, Sanjeewa2020}. However, newer reports of transmission measurements and revisited band structure calculations conflict with these earlier results and propose \ECA\ as a magnetic semi-metal or insulator rather than as a Weyl semi-metal causing some doubt about the material's promise \cite{Shi2024, SantosCottin2023}.

Therefore, additional candidate Weyl Hydrogen atom materials are highly desirable. Recently, \KMAO\ was proposed as such a candidate in a comprehensive \textit{ab initio} study of the compound's electronic properties \cite{Nie2022}. In Ref.~\onlinecite{Nie2022} S. Nie \textit{et al}., performed detailed symmetry analysis and band structure calculations on \KMAO\ which suggest that, for $C_{2z}$ preserving ferromagnetic order, \KMAO\ generates a single pair of Weyl nodes. Furthermore, due to it's unique band structure, they find that it could be a highly tunable topological superconductor with numerous potential topological superconducting states \cite{Nie2022}. Significantly, \KMAO\ is a known compound that has been studied since 2012 for its potential as an ionic conductor and is, therefore, available and accessible to experimental interrogation  \cite{Chaalia2012}. 

\KMAO\ was first reported in Ref.~\onlinecite{Chaalia2012} in a study of its high temperature ionic conductivity and crystal structure. In this work, it was found to crystallize in a monoclinic \Ctc\ symmetry belonging to the highly flexible alluaudite-type family of structures. Central to these materials are transition metal octahedra which form into quasi-1D chains networked by pnictide tetrahedra into a 3D framework with channels that host the alkali (or alkaline) metals leading to a general formula of \textit{X}(2)\textit{X}(1)\textit{M}(1)\textit{M}(2)\textsubscript{2}(\textit{Pn}O\textsubscript{4})\textsubscript{3} (where \textit{X}, \textit{M} and \textit{Pn} are alkali/alkaline metal, transition metal, and pnictide sites perspectively)\cite{Hatert2000,Haj2008}. This structure is highly configurable being adopted by dozens of known compounds, can be pushed into the garnet structure with the suitable choice of stoichiometry, and can even stabilize vacancy structures \cite{Khorari1997}. Interestingly, several of the members of this family have been shown to exhibit magnetic order including ferrimagnetism at 19 K in \AFMPO\ and antiferromagnetism at 35 K in \NCFPO\ demonstrating a tunability to both the ordering temperature and the nature of the ground state \cite{Chouaibi2001,Hidouri2004}. More recently, the alluaudites have shown promise as catalyst and cathode materials due to their channel structure, uniting interests across divergent fields \cite{Buzlukov2020,Kacimi2005,Pan2013,Peng2024}   

\begin{table}
	\caption{\label{tab:one}Crystallographic parameters of \KMAO\ at 293 K.}
	\begin{ruledtabular}
		\begin{tabular}{lllll}
     \multicolumn{2}{l}{Temp.} & \multicolumn{3}{c}{293 K} \\ 
	\multicolumn{2}{l}{Space Group} & \multicolumn{3}{c}{$C2/c$}  \\
	\multicolumn{2}{l}{$R_{wp}$} & \multicolumn{3}{c}{0.066} \\
	\multicolumn{2}{l}{$\chi^2$} & \multicolumn{3}{c}{1.089}  \\
	\multicolumn{2}{l}{$a$ (\AA)} & \multicolumn{3}{c}{12.4354(9)}  \\
    \multicolumn{2}{l}{$b$ (\AA)} & \multicolumn{3}{c}{13.0599(8)}  \\
	\multicolumn{2}{l}{$c$ (\AA)} & \multicolumn{3}{c}{6.8165(4)}  \\
    \multicolumn{2}{l}{$\beta$ (deg)} & \multicolumn{3}{c}{114.234(6)}  \\
	\multicolumn{2}{l}{$V$(\AA$^3$)} & \multicolumn{3}{c}{1009.5(1)}  \\
	\hline
 \multicolumn{1}{l}{Atom (WP) } & \multicolumn{1}{c}{$x$} & \multicolumn{1}{c}{$y$} & \multicolumn{1}{c}{$z$} & \multicolumn{1}{c}{$U (\text{\AA}^2$)} \\
 \hline
\multicolumn{1}{l}{K1 ($4e$)}	& 0	& 0.4857(1)  &	0.25 & 0.04(1)	\\
\multicolumn{1}{l}{K2 ($4a$)}	& 0	& 0  &	0 & 0.058(1)	\\
\multicolumn{1}{l}{Mn1 ($4e$)}	& 0	& 0.7654(1)  &	0.25	& 0.03(1)\\
\multicolumn{1}{l}{Mn2 ($8f$)}	& 0.2768(1)	& 0.1565(1)  &	0.3625(1)	& 0.02(1)\\
\multicolumn{1}{l}{As1 ($4e$)}	& 0	& 0.2175(1)  &	0.25	& 0.011(1)\\
\multicolumn{1}{l}{As2 ($8f$)}	& 0.2373(1)	& 0.3904(1)  &	0.1307(1)	& 0.01(1)\\
\multicolumn{1}{l}{O1 ($8f$)}	& 0.3360(1)	& 0.1672(1)  &	0.1082(1)	& 0.02(1)\\
\multicolumn{1}{l}{O2 ($8f$)}	& 0.0456(1)	& 0.2901(1)  &	0.4718(1)	& 0.02(1)\\
\multicolumn{1}{l}{O3 ($8f$)}	& 0.3814(1)	& 0.4040(1)  &	0.1899(1)	& 0.02(1)\\
\multicolumn{1}{l}{O4 ($8f$)}	& 0.2236(1)	& 0.3184(1)  &	0.3254(1)	& 0.02(1)\\
\multicolumn{1}{l}{O5 ($8f$)}	& 0.3292(1)	& 0.0038(1)  &	0.3812(1)	& 0.03(1)\\
\multicolumn{1}{l}{O6 ($8f$)}	& 0.1000(1)	& 0.1336(1)  &	0.2373(1)	& 0.04(1)\\

		\end{tabular}
	\end{ruledtabular}
\end{table}

In this paper, we revisit \KMAO\ to study its magnetic properties and evaluate its potential as a candidate Weyl Hydrogen atom. Using flux and solid state synthesis we grew high quality single crystal and powder samples which exhibit the structure and stoichiometry reported in Ref.~\onlinecite{Chaalia2012}. Low temperature bulk characterization reveals two magnetic transitions which isothermal magnetization measurements suggest to be to ferrimagnetic states. Interestingly, heat capacity measurements show only a single transition, perhaps indicating the higher temperature signal observed in the magnetization is not to a fully ordered state. Optically, the samples appear brown and transparent belying an insulating behavior which is borne out by a resistivity in excess of 1 M$\Omega$ for a sub cm long sample. Ellipsometry measurements reveal a band-gap of $\approx$ 4 eV, strongly contradicting the predicted semi-metallic behavior. Powder neutron diffraction measurements confirm the two magnetic transitions first to a \k $=(0,0,0)$ ferrimagnetic structure, then to a lower temperature double \k\ structure with \ko $=(0,0,0)$ and \kt $=(1,0,0)$, which also exhibits ferrimagnetism. Notably, both the heat capacity and neutron powder diffraction data indicate the higher temperature magnetic phase may only exhibit relatively short-range correlations as evidenced by broad, weak peaks in the diffraction pattern. Finally, density functional theory calculations are performed which find an antiferromagnetic ground state and a small band gap in contrast to previous reports. Nonetheless the experimental and predicted properties are disparate suggesting a complexity to the electronic correlations not captured by the approach of standard density functional theory. Overall, \KMAO\ exhibits a rare 2-\k\ magnetic structure with evidence of disorder in a highly configurable quasi-1D structure, perhaps offering a rich platform for correlated electron physics in a low-dimensional lattice and routes to drive it closer to the desired magnetic and semi-metallic state.   

\section{\label{sec:Met}Methods}

\subsection{\label{subsec:growth} Materials Synthesis}

The synthesis of \KMAO\ powder samples was performed in air using stoichiometric amounts of the corresponding oxides, carbonates and arsenates. In the reaction, MnO, K$_2$CO$_3$ and NH$_4$H$_2$AsO$_4$ in the molar ratio 3:1:3 were ground thoroughly (for 15 mins) using an agate mortar and pestle. The resulting mixture was loaded into a porcelain crucible and the reaction mixture was heated to 400 \degrees C over 4 hrs, then heated to 800 \degrees C over 10 hrs. The temperature was held at 800 \degrees C for 48 hrs after which the melted mixture was cooled to room temperature at 5 \degrees C/hr. 

Single crystals of \KMAO\ were grown using a KCl flux in a ceramic crucible covered with a lid. The reactants were ground and loaded in a nitrogen-blanket dry box and then heated in a programmable furnace. Crystals were grown by introducing the reactants, MnO (0.4301 g Aldrich, 99+\%), Mn$_2$O$_3$ (0.2393 g, Aldrich, 99\%), NH$_4$H$_2$AsO$_4$ (1.4457 g Thermo Scientific, 98\%) and K$_2$CO$_3$ (0.4189 g, Thermo Scientific, 99\%), to KCl (7.602 g, Aldrich, 99+\%) flux with a flux to charge ratio of 3:1. These were ground thoroughly and the resulting mixture was loaded into a porcelain crucible and the reaction mixture was heated to 400 \degrees C over 4 hr, then 800 \degrees C over 10 hrs. The temperature was held at 800 \degrees C for 48 hrs before the melted mixture was cooled to room temperature at 5 \degrees C/hr. The obtained solid was washed with deionized water using suction filtration methods to eliminate the vitreous mass and isolate \KMAO . The obtained crystal samples (shown in Fig.~\ref{fig:struct} (a)) were small and needle-like with typical dimensions of $\approx 1 x 0.1 x 0.1$ mm. Hundreds of such crystals were produced by each batch. A few larger plate-like crystals of dimensions $\approx 2 x 1 x 0.5$ were also produced. 

\subsection{\label{subsec:exp} Experimental Methods}

Neutron powder diffraction (NPD) experiments were performed on the WAND$^2$ and HB-2A diffractometers of Oak Ridge National Laboratory's High Flux Isotope reactor \cite{Calder2018}. Due to its high flux, WAND$^2$ was used (with incident wavelength 1.48 \AA) to collect a NPD patterns with a fine temperature grid and follow the magnetic order parameter via the peak intensity, to accommodate the low symmetry monoclinic structure an extra 16' collimator was used before the sample. To collect high resolution patterns for magnetic structure solution HB-2A was used with an incident wavelength of 2.41 \AA\ and collimator settings of open-21'-12' for the pre-monochromator, pre-sample, and pre-detector collimators, respectively. The WAND$^2$ experiment used $\approx$ 2 g of powder while the HB-2A experiments used $\approx$ 4 g of powder. For both instruments vandium sample cans were used. We note that a single crystal experiment was performed at WAND$^2$; however, the crystals large enough for neutron diffraction exhibited a high degree of twinning which prevented their use in the magnetic structure solution. Rietveld refinements were performed using the FullProf Suite, and magnetic structure solution was performed using Isodistort, the Bilbao Crystallographic Server, and SARAh \cite{Campbell2006,Aroyo2006,Aroyo2006b,Wills2000,Rodriguez1993}. Crystal structure images where produced using VESTA \cite{Momma2011}  

\begin{figure}[H]
\includegraphics[width=\columnwidth]{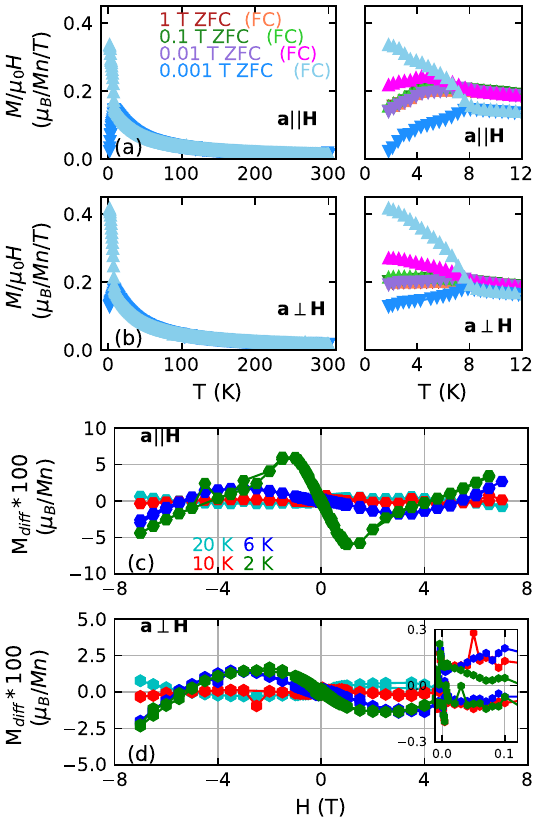}
\caption{\label{fig:mag}Bulk magnetic susceptibility measured with the applied field (a) along the \textbf{\textit{a}} axis and (b) perpendicular to the \textit{\textbf{a}}-axis. Data collected using field cooled and zero-field cooled procedures are shown as up and down pointing triangle markers, respectively. Field isotherms collected for the same (c) parallel and (d) perpendicular directions with the linear component of the $M(H)$ curve subtracted off. In panels (c) and (d) the signal has been multiplied by 100. The inset of (d)  shows a zoomed in view near the origin showing a small hysteresis. }
\end{figure}

Single crystal x-ray diffraction (XRD) data were collected using a Rigaku XtalLAB Synergy S single crystal diffractometer with a Mo x-ray source. Several crystals across numerous synthesis batches were measured by mounting them to small kapton loops with a minimal amount of oil. Data collection planning, data reduction, and structure solution was performed using the CrysAlis\textsuperscript{Pro} software. Structure refinements were performed using Shelxl \cite{Sheldrick2015}

Bulk properties were measured on multiple single crystal and powder samples. The data shown in this work was all collected on a single sample across the different measurements and is representative of all the single crystal measurements. We note that magnetic impurities in the powder sample contaminated the bulk probe measurements with additional transitions and so only the single crystal data is used in the analysis. Measurements of the heat capacity and attempted measurements of the resistivity were performed using a Quantum Design Physical Properties Measurement System. For the heat capacity measurements addenda were measured and used to process the raw data. Magnetization measurements were performed using a Quantum Design Magnetic Property Measurement System 3. The crystals were attached to quartz rods with grease and measurements were collected under field and zero-field cooled procedures. 

Variable angle spectroscopic ellipsometry measurements were performed using a Woollam M-2000 ellipsometer. A sample of \KMAO was prepared by mounting a regularly shaped single crystal to a copper plate and polishing it to a smooth surface with platelet like dimensions of $\approx 1\times1$mm. Data modeling was performed using Woollam CompleteEASE software using a General Oscillator model with Tauc-Lorentz oscillators. 

\subsection{\label{subsec:DFT} First Principle Calculations}

Density functional theory calculations were carried out using the linearized augmented plane-wave density functional theory code WIEN2K \cite{Blaha2001}. All calculations initially employed the generalized gradient approximation of Perdew \textit{et. al.}, with a U of 5 eV added to the Mn orbitals to better depict this material \cite{Perdew1996,Anisimov1997}. The structure used was that reported by Chaalia \textit{et. al.} and confirmed in this work. Internal coordinates were relaxed within an assumed FM configuration, to guard against possible magneto-elastic effects [3-12], and the structure employed for all other calculations. We checked that the forces in the other magnetic configurations, with this physical structure, are reasonably small (under 6 mRyd-bohr vs. under 2 mRyd-bohr in the ferromagnetic configuration). Spin-orbit coupling was included, with the magnetization axis chosen as [001].

\section{\label{sec:Struct}Crystal Structure}

The crystal structure of \KMAO , as determined using single crystal XRD, is shown in Fig~\ref{fig:struct}. Our samples crystallize in the alluaudite structure with monoclinic space group symmetry $C2/c$ (see Table~\ref{tab:one}) as previously reported \cite{Chaalia2012}. The structure is composed of MnO$_6$ octahedra which assemble into infinite chains along the $[10\overline{1}]$ direction. These chains are capped by AsO$_4$ tetrahedra which join neighboring MnO chains creating layers in the \textbf{\textit{ac}} plane. The AsO$_4$ tetrahedra also link neighboring chains along the \textit{\textbf{b}} axis, creating channels along the \textit{\textbf{c}} axis which are filled by the K$^+$ ion.     

The crystal structure has two symmetry distinct K, Mn, and As sites and six distinct O  sites in the unit cell defining a relatively low symmetry structure. In particular the two Mn sites (Mn1 and Mn2) have different multiplicities and experience different local octahedral O coordinations potentially leading to distinct magnetic behavior on either site. The $8f$ Mn2 site is less distorted, having O-Mn-O angles between 77.5 and 110.0 degrees while the $4e$ Mn1 site varies between 71.9 and 117.7 degrees.  The MnO$_6$ chains are constructed of alternating sites with a Mn1-Mn2-Mn2-Mn1-Mn2-Mn2 sequence. The low symmetry of the $8f$ site in this sequence causes a zig-zag structure to the chain roughly in the plane of the chain direction and the \textit{\textbf{b}} axis, with smaller deviations in the other directions which, when projected along the chain direction, leads to the appearance of an \lq $X$\rq\ stretched along the \textit{\textbf{b}} direction.  Each unit cell contains four such chains which are symmetry equivalent and are situated as pairs on either side of the $\mathbf{b} = \frac{1}{2}$ plane.

\section{\label{sec:bulk}Bulk Magnetic Properties} 

Fig.~\ref{fig:mag} (a) and (b) show the magnetic susceptibility of \KMAO\ collected both parallel and perpendicular to the growth direction (i.e. the \textit{\textbf{a}} axis) under $\mu_0 H = 1$ mT. Qualitatively, both $M(T)$ curves look similar with typical paramagnetic behavior down to approximately 10 K and no obvious anisotropy in the PM state. Using standard Curie-Weiss fitting results in Curie-Weiss temperatures ($\theta_CW$) and effective moments ($\mu_{eff}$) of -16 K, 5.9 $\mu_B$ and -17 K, 6.5 $\mu_B$ for the parallel and perpendicular directions, respectively. The negative $\theta_CW$ values indicate AFM interactions while the $\mu_{eff}$ are generally consistent with Mn$^{2+}$ in a high spin state, although they are a bit higher than that expected for Mn$^{3+}$. For the zero field cooled (ZFC) data, below $\approx 8$ K the curves diverge from Curie-Weiss behavior, exhibiting sharp downturns indicative of magnetic ordering. At $\approx 4$ K another slight kink in the susceptibilities is observed, potentially indicating a second transition. For the data collected under field-cooling (FC), at 8 K a large increase in the signal is seen which then peaks and exhibits a kink with further increases at the second transition at 4 K, indicating the 8 K transition is likely either ferro- or ferrimagnetic. Comparing the two field directions, the response is stronger perpendicular to the \textit{\textbf{a}}-axis by $\approx$ 20 $\%$ suggesting a larger ferromagnetic moment component perpendicular to \textit{\textbf{a}}. With stronger probe fields two changes are observed: the FC/ZFC splitting decreases, and the second transition exhibits a clear downturn in the FC data for \textit{\textbf{a}} $\parallel $ \textbf{H}. 

We next consider the $M(\textbf{H})$ isotherms (Fig.~\ref{fig:mag} (c) and (d)). Unexpectedly given the field dependent behavior observed in the $M(T)$ curves, in the raw isotherms we do not see significant hysteresis or any evidence of saturation up to the maximum measured field of 7 T at which point the magnetization only reaches 1.5 $\mu_B$/Mn - much less than the expected saturated Mn$^{2/3+}$ moment. Therefore, to give better insights to the magnetization in Fig.~\ref{fig:mag} (c) and (d) we subtract off the linear component using linear fits across the full measured range for each temperature and field orientation (see the supplemental materials for example fits) \cite{SM}.  

\begin{figure}
\includegraphics[width=\columnwidth]{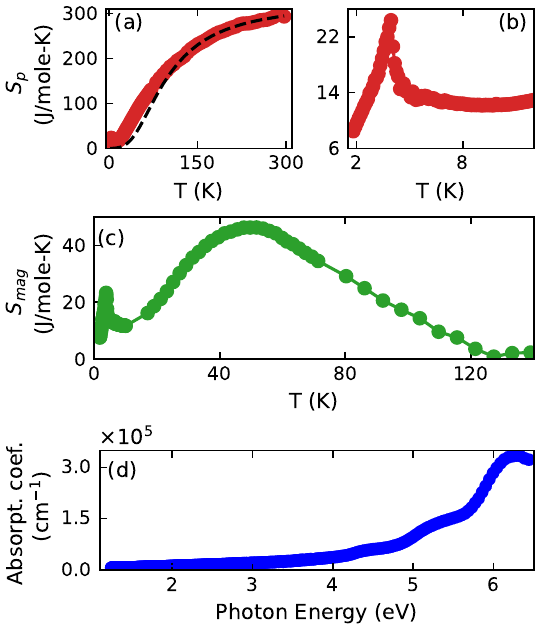}
\caption{\label{fig:HC} (a) Zero Field heat capacity of a single crystal (same sample used in the magnetization measurements) of \KMAO . The data are shown in red markers while a Debye model fit to the phonon contribution is shown as a dashed line. (b) A higher point density measurement of the low temperature heat capacity. (c) Magnetic heat capacity, as determined by subtracting the phonon contribution from the Debye model fitting.  (d) Extracted absorption constant for \KMAO\ as a function of incident photon energy.}
\end{figure}
 
In the 20 K difference curve ($M_{diff}(\textbf{H})$) we see a slight enhancement of the magnetization for low fields for both orientations, indicating the presence of some polarizable unordered moments. At 10 K and below (also for both orientations) this behavior changes to a negative slope where the sample's magnetization lags that of the linear fit. Looking closely at the low field region, we see a slight hysteresis at all temperatures which closes by $\approx$ 1 T - consistent with the $M(T)$ field dependence. However, at 6 K where we expect to be below the first magnetic transition, the curve looks very similar to that of 10 K with no obvious hysteresis or saturation. Below the 6 K feature, the negative slope increases slightly in magnitude for the parallel configuration. For the  applied field parallel to \textit{\textbf{a}} the response becomes stronger with a peak at 1.5 T. These differences in the $M(\textbf{H})$ curves possibly suggest a local environment anisotopy with a preference for the moment to point in a direction $\perp$ \textit{\textbf{a}} with a strong easy direction that prevents the moments from aligning with the applied field along either \textbf{\textit{a}} or in the \textit{\textbf{bc}} plane for fields up to 7 T. The slight non-linear behavior with a small hysteresis seen below 10 K is indicative of a soft ferromagnet or of a ferrimagnet.  

Fig.~\ref{fig:HC}(a) and (b) show the heat capacity measurements of the same single crystal sample used in the magnetization measurements. Typical temperature dependence is observed until $\approx$ 8 K where the heat capacity has an inflection point which at lower temperatures develops into a $\lambda$ anomaly peaked at $\approx$ 4 K. While the 4 K transition agrees well with the second transition observed in the $M(T)$ curves, there is no obvious sign of the 8 K transition. This may be due to a smaller signal associated with only part of the magnetic sublattice ordering and the proximity of the second transition to a fully ordered state which would have a larger change in the heat capacity, such a scenario will be discussed later in this work. To isolate the magnetic contribution to the heat capacity, the high temperature portion of $S_p$ was fit using the Debye model, while not ideal for this temperature range this model was used as no nonmagnetic analog material was available to experimentally determine the lattice contribution. The thus obtained bare magnetic heat capacity is plotted in Fig.~\ref{fig:HC}(c) and reveals a large broad feature between 120 and 8 K in addition to the previously discussed lower temperature transition. Such a feature is common in low dimensional magnetic materials with short range magnetic correlations \cite{Ortiz2023,Xing2019,Ranjith2019}.


Finally, we note that given the predicted semi-metallic band structure of \KMAO\ in ref.~\cite{Nie2022}, attempts were made to measure the resistivity of a single crystal. However, the as grown crystals were optically transparent and all attempts to measure resistivity using standard six probe procedures were unsuccessful indicating a resistance of greater than 1M$\Omega \ mm$.  This is consistent with the conductivity measurements reported by Chaalia \textit{et al}., in ref.~\cite{Chaalia2012} which used complex impedance spectroscopy and found a high temperature conductivity of $3.6\times 10^{-8} \mathrm{ \Omega}^{-1}\mathrm{cm}^{-1}$.  To better quantify the insulating state ellipsometry measurements were performed on a polished single crystal sample (Fig.~\ref{fig:HC}(d)). The absorption spectra reveals a large multi-eV bandgap with little absorption observed for photons of energy less than 4 eV. This is largely at odds with suggestions of a semi-metallic state but consistent with the observed optical transparency. From these measurements it must be concluded that \KMAO\ is an insulator.

\section{\label{sec:neutron}Neutron Scattering and the Magnetic Structure} 

\begin{figure}[h]
\includegraphics[width=\columnwidth]{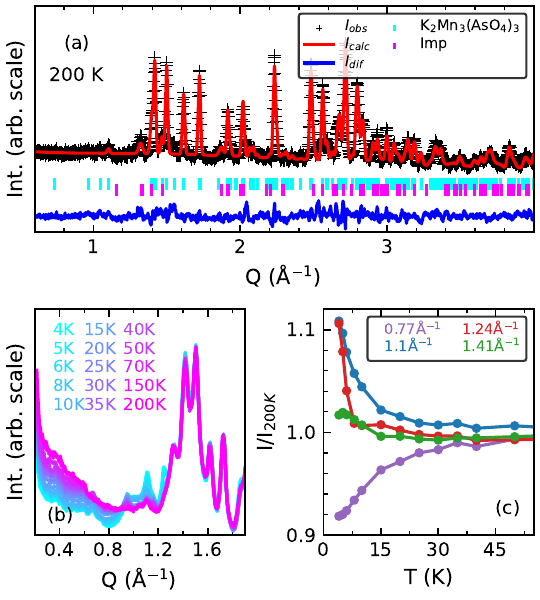}
\caption{\label{fig:diffr} (a) Neutron powder diffraction pattern of \KMAO\ collected at 30 K on HB-2A shown together with the calculated pattern from a best fit structural model determined from Rietveld refinements. (b) Plot of the low-$Q$ region of neutron powder diffraction data collected on WAND$^2$ for temperatures between 200 and 4 K. (c) Temperature dependence of various regions of panel (b) plotted as intensities integrated over a $\Delta Q = 0.3$ \iA\ window. }
\end{figure}

To determine the microscopic magnetic properties of \KMAO , NPD experiments were performed. Fig.~\ref{fig:diffr} (a) shows a NPD pattern collected at 200 K (in the paramagnetic state) on HB-2A together with a calculated best fit pattern using a purely nuclear structural model. As seen, the structural model determined from single crystal XRD accounts fairly well for the observed intensity, though we note several small unindexed peaks. Most of these could be indexed with small ($<$ 5\%\ volume fraction) impurity phases of MnO$_2$ and Mn$_3$As, as both of these phases have magnetic transitions of their own, their respective magnetic structures were also included in the refinements using the structures reported on the Bilbao Crystallographic server \cite{Gallego2016, Brown1990, Sato2001}. 

\begin{figure}
\includegraphics[width=\columnwidth]{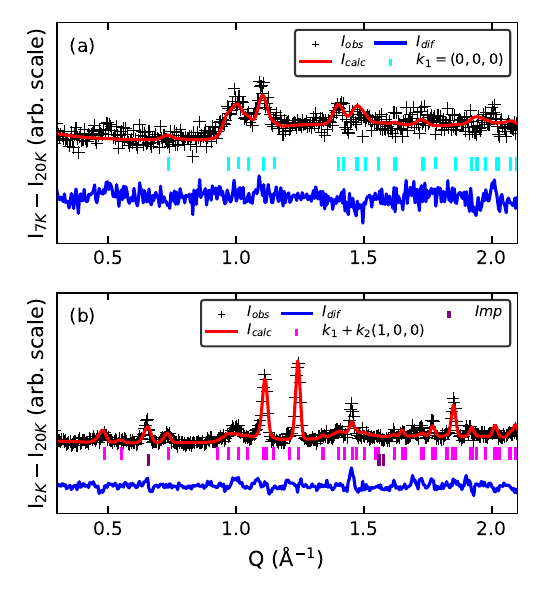}
\caption{\label{fig:magsol} Difference curves of NPD data for (a) 15 minus 7 K and (b) 15 minus 2 K to isolate the magnetic signal in the diffraction patterns. Also plotted are the results of Rietveld refinements using the magnetic structures discussed in the main text. }
\end{figure}

Fig.~\ref{fig:diffr} (b) shows a plot of NPD patterns collected between 200 and 4 K on WAND$^2$ focusing on the low $Q$ region. Here, several new peaks emerge as the temperature is decreased below $\approx$ 15 K, most notably around $Q \approx 1.21 \text{ \AA}^{-1}$. Additionally, several nuclear peaks increase in intensity (such as the peak at 1.41 \iA ). These two classes of peaks indicate ordering vectors (\k) of \k $\neq 0$ and \k $=0$ respectively. Also observable is a rather significant decrease in a diffuse low-$Q$ ($Q < 0.9 \text{ \AA}^{-1}$) scattering signal upon cooling.  Interestingly, this diffuse signal appears to have some structure (as seen by the anisotropic $Q$ dependence such as the dip in intensity at 0.8 \iA). This is consistent with short-range magnetic correlations and the broad diffuse feature in the magnetic heat capacity. The shift in spectral weight from diffuse to Bragg reflections is indicative of long-range order setting in over a fraction of the locally correlated spins.

Fig.~\ref{fig:diffr}(c) shows the temperature dependence of the integrated intensity (over a $\Delta Q = 0.04$ \iA\ range and normalized to the 200 K value) for several of the observed peaks and for the low-$Q$ background. For the latter (here taken at $Q = 0.77$ \iA ) the intensity decreases around 40 K and continues to do so down to 4 K with an inflection point around the $\approx$ 8 K transition, consistent with it arising from PM scattering of unordered Mn moments. Considering the structural peaks with increased intensity at 1.1 and 1.41 \iA , we observe a gradual increase of intensity below $\approx$ 10 K at both peak positions. For the peak at 1.1 \iA\ this increase continues down to 4 K while for the 1.41 \iA\ peak it at first increases and then decreases below $\approx$ 5 K. The peak at 1.24 \iA\ does not correspond to a structurally allowed peak position and shows a steep increase in intensity below $\approx$ 6 K.  Using these temperature dependencies, we can associate these peaks with the transitions observed in the magnetization data. The 1.41 \iA\ peak should result from the first 8 K transition to a $\textbf{k}=0$ order while the 1.24 \iA\ peak arises from the second 4 K transition to $\textbf{k} \neq 0$ order. Given its non-monotonic behavior, the peak at 1.1 \iA\ appears to have contributions from both magnetic phases. We note that while the second transition appears sharp in the neutron scattering data (as seen by the rapid increase in the 1.24 \iA\ peak's intensity), the 1.1 and 1.41 \iA\ peaks exhibit a broader transition. Intensity at these positions slowly increases starting at $\approx$ 25 K - a similar temperature at which the background signal begins decreasing - perhaps indicating the onset of short range magnetic correlations related to this phase prior to the onset of long-range magnetic order. It should be noted that given the lack of energy discrimination in the diffraction data, we are unable to discern between fluctuating and static magnetic correlations.   

From the above discussion we can identify a magnetic ordering vector \ko\ $ = (0,0,0)$ for the 8 K transition. Using this $\textbf{k}$, potential magnetic structures were generated using representational analysis (as shown in Tab.~\ref{tab:two}). For \ko , \Ctc , and Wyckoff positions $8f$ and $4e$, four irreducible representations (irreps $\Gamma$) are generated. All four irreps are allowed for both Mn sites, with $\Gamma^{k_1}_1$ and $\Gamma^{k_1}_2$ being of higher symmetry with four total basis vectors and $\Gamma^{k_1}_3$ and $\Gamma^{k_1}_4$ consisting of five total basis vectors. Given the multiple Mn sites per unit cell, the $\textbf{k}=0$ ordering vector is able to generate AFM orders and all irreps have both FM and AFM basis vectors. 

To solve the magnetic structure, all four irreps were trialed in Rietveld refinements against the NPD data. The main phase magnetic signal was isolated through the use of NPD difference curves, subtracting the 20 K pattern from the 7 K and 2 K patterns. 20 K was used as the background pattern, rather than 10 K, as diffuse magnetic peaks were observed even at 15 K (see Fig.~\ref{fig:diffr}). Similarly 7 K was used rather than a lower temperature to prevent from contaminating the first transition fits with magnetic intensity from the lower transition. In general, in the NPD the magnetic transitions seem to occur at slightly higher temperatures than observed in the bulk probe measurements likely due to the difference in sample size and between the thermal conductivity of a single crystal vs. a multi gram powder sample and the data collection occurring on warming. 

\begin{figure}
\includegraphics[width=\columnwidth]{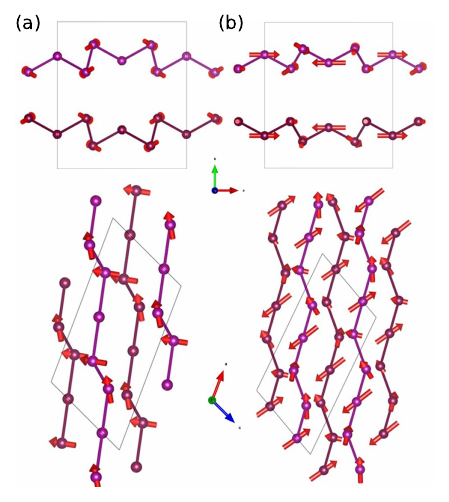}
\caption{\label{fig:magstruct}Magnetic structure of \KMAO\ shown (a) for the \ko\ structure and (b) the \ko $+$ \kt\ structure.  Mn chains on the $y\approx0.25$ plane are shown in burgundy while those on the $y\approx 0.75$ plane are shown in purple. To aid in visualization, the vectors representing the magnetic moments have different scale factors for the two structures and therefore do not represent the relative moment size.}
\end{figure}

Fig.~\ref{fig:magsol} (a) shows the best fit for the 7 K data which was achieved using a combination of the $\Gamma^{k_1}_3$ and $\Gamma^{k_1}_4$ irreps, the resulting magnetic structure is shown in Fig.~\ref{fig:magstruct} (a). In this model, the basis vectors corresponding to moments on the Mn2 site were found to be most vital to obtaining a decent fit and dictated the need for two irreps. The use of either only $\Gamma^{k_1}_3$ or $\Gamma^{k_1}_4$ led to fits which only adequately modeled either the peak at $\approx 1.0$ or $\approx 1.1$ \iA\ and only their combination $\Gamma^{k_1}_3 \oplus \Gamma^{k_1}_4$ fit both of these peaks without contributing additional intensity at lower $Q$ where no peaks were observed experimentally. It is somewhat unusual to have a single transition invoking two irreps which should indicate multiple order parameters and a transition with a first order character. 

The resultant magnetic structure (Fig.~\ref{fig:magstruct}) has an alternating arrangement of moments within each chain. The nearest neighbor Mn2-Mn2 order is FM along \textit{\textbf{c}} but AFM along \textit{\textbf{a}} and \textit{\textbf{b}}. Across the Mn1 site the Mn2 moments are FM in \textit{\textbf{a}} and \textit{\textbf{c}} but AFM along \textit{\textbf{b}}. All of the Mn2 sites share the same moment size with a refined magnitude of $\approx 2.0 \mu_B$. The small ($\approx 0.5 \mu_B$) moment obtained on the Mn1 sites in this model roughly aligns ferromagnetically with its neighboring Mn2 sites however, it's moment is always largely along the Mn chain direction. This structure corresponds to the magnetic space group (MSG) $C2'$ \cite{Perez2015}. We note that given the quality of the data (which exhibits relatively few, broad peaks) and the low symmetry of the structure that, while this model provides the best fit, the details of this structure require higher quality, preferably single crystal data to study carefully.  We also checked these models against both the HB-2A and WAND$^2$ data sets and each resulted in the same solution.  

Given these magnetic peaks are quite broad and weak, the lack of a discrete transition temperature in the temperature dependence of the peak intensity, and the missing signal in the heat capacity for the 8 K transition it is possible that the magnetism observed at 7 K is only partially ordered. Considering the quasi-1D structure to the Mn sublattice, the symmetry distinct Mn sites and the complex local environments, there are multiple potential causes for magnetic frustration and disorder in the material. Magnetic disorder in quasi-1D chains is relatively common, especially in low symmetry compounds with numerous potential competing interactions along the various exchange pathways \cite{Hohenberg1967, Mermin1966, Ramirez1994,Zaliznyak1999,Lake2000}. However, this could also be due to the somewhat limited temperature range between the two transitions which may lead to the second order onsetting before the first transition has fully saturated.  Additional inelastic neutron scattering experiments to look at the spin excitations would be interesting should higher purity powder samples become available.

\begin{table}
	\caption{\label{tab:two} Irreducible representations ($\Gamma$), number of basis vectors ($\psi$), magnetic space group (MSG), magnetic super cell, and origin shift for $\textbf{k} = (0,0,0)$. Separate $\Gamma$ and $\psi$ are generated for either Mn site. The Mn1 site's $\psi$ have two components corresponding to the two symmetry generated Mn positions (0, 0.7654, 0.25) and (0, 0.2346, 0.75).  Similarly the Mn2 site's $\psi$ have 4 components for the four generated positions (0.2768, 0.1565, 0.3625), (0.7768, 0.6565, 0.3625), (0.7232, 0.1565, 0.2375), (0.7232, 0.8435, 0.6375), and (0.2768, 8435, 0.8625). }
	\begin{ruledtabular}
		\begin{tabular}{cclc}
    		 \multicolumn{1}{c}{$\Gamma$}  & \multicolumn{1}{c}{$\psi$} & \multicolumn{1}{c}{Vector components} & \multicolumn{1}{c}{MSG}  \\
	\hline
	\multirow[t]{2}{*}{$\Gamma^{k_1}_1\ $}  &  1  &  $b:(++)$ & \multirow[t]{2}{*}{$C2/c$}     \\
									 &  3  &  $a:(+-+-)$ &           \\
          						 &     &  $b:(++++)$ &           \\
                    				 &     &  $c:(+-+-)$ &           \\
	\multirow[t]{2}{*}{$\Gamma^{k_1}_2\ $}  &  1  &  $b:(+-)$   & \multirow[t]{2}{*}{$C2/c'$}    \\
									 &  3  &  $a:(+--+)$ &           \\
          						 &     &  $b:(++--)$ &          \\
                    				 &     &  $c:(+--+)$ &           \\
	\multirow[t]{2}{*}{$\Gamma^{k_1}_3\ $}  &  2  &  $a:(++)$   & \multirow[t]{2}{*}{$C2'/c'$}    \\
									 &     &  $c:(++)$ &           \\
									 &  3  &  $a:(++++)$ &           \\
          						 &     &  $b:(+-+-)$ &           \\
                    				 &     &  $c:(++++)$ &           \\
	\multirow[t]{2}{*}{$\Gamma^{k_1}_4\ $}  &  2  &  $a:(+-)$ &\multirow[t]{2}{*}{$C2'/c$}     \\
									 &     &  $c:(+-)$ &           \\
									 &  3  &  $a:(++--)$ &          \\
          						 &     &  $b:(+--+)$ &           \\
                    				 &     &  $c:(++--)$ &           \\ 

		\end{tabular}
	\end{ruledtabular}
\end{table}

Moving to the base temperature magnetic structure, the \ko\ ordering vector was insufficient to index all of the magnetic reflections and, indeed, no single $\mathbf{k}$ was found to adequately index the obtained pattern. Therefore, to solve the structure a second $\mathbf{k}$ vector was introduced and determined to be \kt\ $ = (1,0,0)$. This $\mathbf{k}$ breaks the $C$ centering symmetry of the structural space group which is preserved in the \ko\ structure. The results of representational analysis and the corresponding MSG for this $\mathbf{k}$ in isolation are shown in Tab.~\ref{tab:three}. As before, \kt\ results in four irreps with the same splitting between two higher symmetry irreps with four basis vectors each and two lower symmetry irreps with five basis vectors. Of these, $\Gamma^{k_2}_3$ (in combination with $\Gamma^{k_1}_3 \oplus \Gamma^{k_1}_4$) which together correspond to the MSG $P2'$, was found to produce the best fit to the observed difference pattern (Fig.~\ref{fig:magsol} (b)).  

To achieve a quality fit, both irreps were required on both magnetic sites, leading to the use of ten basis vectors in total. Given, that for the 8 K structure the Mn2 site had a significantly larger ordered moment, attempts were made to model the data with the two $\mathbf{k}$ vectors isolated to the different Mn sublattice, with \ko\ on Mn2 and \kt\ on Mn1. However, such models did not produce convincing fits and so both sets of basis vectors were allowed on either site. The resulting best fit magnetic model is shown in Fig.~\ref{fig:magstruct}(b). Here, the Mn1 moments form FM stripes along the \textit{\textbf{c}} axis which align AFM along the \textit{\textbf{a}} direction. These larger Mn1 moments form a $\approx$ 30\degrees\ angle with the Mn chain. The broken $C$ centering leads the Mn1 moments related by the centering operations to have $\approx$ 10 \%\ modulation in their amplitudes and a slight $\approx$ 3\degrees\ shift in their tilt in the \textit{\textbf{ac}} plane. From the 8 K transition, the Mn2 moments in general rotate to both be more in the \textit{\textbf{ac}} plane and align with the Mn chain axis. 

The superposition of the two structures leads to  ferrimagnetism even on the individual sublattices with both exhibiting a net moment along the \textit{\textbf{a}} axis and the Mn2 site being largely FM along \textit{\textbf{c}}. The Mn1 sites have an average moment size of $5.5(5) \mu_B$ while the more varied Mn2 site averages $3.6(5) \mu_B$ this is roughly consistent with the sites having Mn$^{2+}$ and Mn$^{3+}$ respectively. We note that while this structure was the best fit found for this study, given the data were collected on a powder sample and the number of basis vectors implemented the specifics of the structure should be taken with some skepticism. Although the $\mathbf{k}$ vectors were uniquely determined and the best fitting irreps clear, specific models within the parameter space of  $\Gamma^{k_2}_3 \oplus (\Gamma^{k_1}_3 \oplus \Gamma^{k_1}_4)$ were dependent on the fit procedure, the modeling of the nuclear peak shape, the Mn site atomic positions and the background. Visually better fits were achieved through the use of simulated annealing and a $P\overline{1}$ structure with spherical moments however, these solutions were not unique and required more parameters than could be justified by the data. This magnetic structure likely requires single crystal data for a robust solution. 

Considering both the 7 K and 2 K magnetic structures there is some general agreement with the previously discussed bulk magnetization measurements. In both structures the \textit{\textbf{a}} lattice direction has largely compensating AFM correlations whereas both \textit{\textbf{b}} and \textit{\textbf{c}} are more strongly ferrimagnetic, consistent with the directional dependence of the measured susceptibility. The ZFC v FC splitting is expected for a FM or ferrimagnetic structure as seen here and the lack of a saturating moment in the isothermal magnetization curves can be explained by the ferrimagnetism, orientation of the magnetic moments, and, more speculatively, the assumption of a large crystal field anisotropy. 

While the specifics of these reported magnetic structures may need additional work to solve, from the \k\ identification and symmetry analysis it is clear that both structures break the desired $C_{2z}$ symmetry needed to generate the predicted Weyl band structure. However, given the insulating nature of the material, this is perhaps a secondary concern. It is notable that a simple polarized FM structure is allowed within the irreps of the \ko\ symmetry table which may be of interest if \KMAO\ can be induced to have a less insulating band structure via doping, chemical pressure, or external pressure. Such changes might be possible via chemical substitution on the \textit{A} site, electron doping with a second transition metal, or even the controlled introduction of vacancies. These approaches could give experimental control not only of the band structure but also of the magnetic ground state, including the spin states. The reported ground state structure is an interesting example of a 2-\k\ magnetic state in a material with a 1D magnetic sublattice which is somewhat rare\cite{Batista2016,Allred2016}. In general quasi-1D magnetic chain materials are of interest in the study of low dimensional physics \cite{Carmelo2023,Coldea2010,Sanjeewa2016,Liu2020,Bao2022,Whitt2022,Taddei2019,Taddei2023,Taddei2018}. Given the flexibility of the alluaudite structure, the spin state, transition temperature and symmetry of the magnetic state are all tunable making them a potential playground for magnetism on a quasi-1D lattice.

\begin{table}
	\caption{\label{tab:three} Irreducible representations ($\Gamma$), number of basis vectors ($\psi$), vector components, and magnetic space group (MSG) for $\textbf{k} = (1,0,0)$. Separate $\Gamma$ and $\psi$ are generated for either Mn site. The Mn1 site's $\psi$ have two components corresponding to the two symmetry generated Mn positions (0, 0.7654, 0.25) and (0, 0.2346, 0.75).  Similarly the Mn2 site's $\psi$ have 4 components for the four generated positions (0.2768, 0.1565, 0.3625), (0.7768, 0.6565, 0.3625), (0.7232, 0.1565, 0.2375), (0.7232, 0.8435, 0.6375), and (0.2768, 8435, 0.8625). }
	\begin{ruledtabular}
		\begin{tabular}{cclc}
    		 \multicolumn{1}{c}{$\Gamma$}  & \multicolumn{1}{c}{$\psi$} & \multicolumn{1}{c}{Vector components} & \multicolumn{1}{c}{MSG/supercell/origin}  \\
	\hline
	\multirow[t]{2}{*}{$\Gamma^{k_2}_1\ $}  &  1  &  $b:(++)$   &  \multirow[t]{2}{*}{$P_C2/c$}     \\
									 &  3  &  $a:(++-+)$ &  \multirow[t]{2}{*}{$(a,b,c)$}         \\
          						 &     &  $b:(+---)$ &  \multirow[t]{2}{*}{$(0,0,0)$}         \\
                    				 &     &  $c:(++-+)$ &           \\
	\multirow[t]{2}{*}{$\Gamma^{k_2}_2\ $}  &  1  &  $b:(+-)$   &  \multirow[t]{2}{*}{$P_C2/c$}    \\
									 &  3  &  $a:(+++-)$ &  \multirow[t]{2}{*}{$(a-2c,-b,-a+c)$}         \\
          						 &     &  $b:(+-++)$ &  \multirow[t]{2}{*}{$(\frac{1}{4},\frac{1}{4},0)$}        \\
                    				 &     &  $c:(+++-)$ &           \\
	\multirow[t]{2}{*}{$\Gamma^{k_2}_3\ $}  &  2  &  $a:(++)$   &  \multirow[t]{2}{*}{$P_C2_1/c$}    \\
									 &     &  $c:(++)$   &  \multirow[t]{2}{*}{$(a-2c,-b,-a+c)$}          \\
									 &  3  &  $a:(+---)$ &  \multirow[t]{2}{*}{$(0,0,0)$}            \\
          						 &     &  $b:(++-+)$ &           \\
                    				 &     &  $c:(+---)$ &           \\
	\multirow[t]{2}{*}{$\Gamma^{k_2}_4\ $}  &  2  &  $a:(+-)$   & \multirow[t]{2}{*}{$P_C2_1/c$}     \\
									 &     &  $c:(+-)$   & \multirow[t]{2}{*}{$(a,b,c)$}          \\
									 &  3  &  $a:(+-++)$ &  \multirow[t]{2}{*}{$(\frac{1}{4},\frac{1}{4},0)$}        \\
          						 &     &  $b:(+++-)$ &           \\
                    				 &     &  $c:(+-++)$ &           \\

		\end{tabular}
	\end{ruledtabular}
\end{table}

\section{\label{sec:DFT} First Principles Calculations}

In an effort to describe the observed magnetic behavior of this rather complex compound, and in particular the large discrepancy between our experimental findings and the previous theoretical work of Ref.~\onlinecite{Nie2022}, we conducted first principles calculations of the magnetic and electronic structure and associated exchange interactions. In view of the large unit cell, these calculations were computationally expensive, and so we limited ourselves to three specific magnetic states within the single crystallographic unit cell (i.e. $Q=0$ states only): the FM posited as the ground state in \onlinecite{Nie2022}, a ferrimagnetic state (FI1), with the Mn1 and Mn2 sites initialized as opposite in orientation, and an effectively AFM state (AF1) with the Mn chains along the a-axis anti-aligned. While none of these states attains the full complexity of either of the magnetic states observed experimentally, the results detailed below do offer some insight into the behavior of this material.  In view of the expense of these calculations, we did not study magnetic anisotropy or non-collinear states.

For all calculations, Mn atomic moments are similar – typically 4.1 to 4.4 $\mu_{B}/$Mn - to the average moment size observed in the neutron diffraction Tab.~\ref{tab:four} below lists the magnetic moments and associated energetics for these states, for the GGA and GGA+U calculations. One notes immediately that the moment magnitude is essentially independent of the magnetic state studied, so that the magnetism in this compound is clearly of a local variety, rather than itinerant.  There is a slightly larger moment on Mn1, at 4.41 $\mu_{B}$ for all states, than the approximate 4.14 value for Mn2, reflecting the significant difference in surrounding environment between these two atomic sites and consistent with the neutron diffraction results.

\begin{table}
	\caption{\label{tab:four} The first-principles-calculated magnetic properties of the magnetic states described in the text. The ground state is indicated by the zero energy, with all other energies relative to this state.}
	\begin{ruledtabular}
		\begin{tabular}{p{0.2\linewidth}  p{0.2\linewidth} p{0.2\linewidth}  p{0.2\linewidth} p{0.2\linewidth}}
    		 State  & GGA Energy (meV/Mn) & GGA+U Mn1 ($\mu_B$) & GGA+U Mn2 ($\mu_B$) & GGA+U Energy (meV/Mn)  \\
	\hline
	   FM   &   14.53   &   4.41         &   4.15        &   11.20    \\
	   FI1  &   0       &   4.41         &  -4.15        &   12.96    \\
	   AF1  &   12.59   &   $\pm$ 4.41   &   $\pm$4.14   &   0    \\

		\end{tabular}
	\end{ruledtabular}
\end{table}


It is also apparent that the ground state (within this particular manifold) depends on the approximation, with the ground state in the unadorned GGA being a ferrimagnetic one and the corresponding state in the GGA+U being an antiferromagnetic one.  Given the known challenges that magnetic transition metal oxides present to the straight GGA, it is our contention that the GGA+U approach is more likely to be appropriate here \cite{Anisimov1997}.

One notes also comparatively large energy differences – greater than 10 meV/Mn – between the calculated ground state and the nearest excited state, which would ordinarily indicate an ordering point on the order of 40 K. This is to be compared with the experimental ordering point of under 10 K, and this, supports our contention that fluctuations or other correlative effects are weakening the magnetism from what a mean-field-based approach would suggest.

Notably, unlike the previous theoretical study, our AFM groundstate band structure does exhibit a small band-gap of approximately 0.35 eV, so that theoretically we at least do not find metallic behavior, although the calculated band gap is roughly an order of magnitude smaller than the highly insulating one found in experiment. We present the calculated band structure below in Fig.~\ref{fig:DFT}. Provocatively, at $\approx 0.1$ eV above the $E_F$, our calculated band structure exhibits a large gap in the band structure. It is possible that a small deviation from the expected stoichiometry, via a small ($\approx$ 1\%) concentration of vacancies for instance, could shift the $E_F$ into this gap which would be more consistent with our ellisometry measurements than the 0.35 eV gap. Given the quality of our NPD and the complexity of the structure, our Rietveld refinements are somewhat agnostic to such a deviation in occupancies for both the O and K sites. Therefore, it is possible that \KMAO\ tends to form with a certain vacancy concentration leading to the observed physical properties.   

\begin{figure}
\includegraphics[width=\columnwidth]{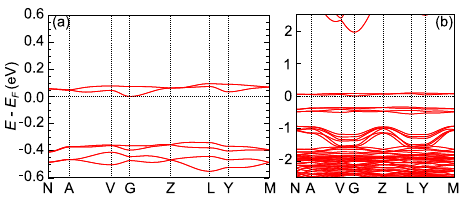}
\caption{\label{fig:DFT}Calculated electronic band structure of \KMAO\ in the described simplified AFM ground state for (a) zoomed in region around the Fermi Energy and (b) large energy range. Note that by convention, the conduction band is set to $E_F$.}
\end{figure}

It is, perhaps, strange that our theoretical work both overstates the strength of the (antiferro)magnetism, a behavior that in oxides is often associated with insulating character, and yet substantially understates the insulating character itself, based on the calculated and measured band gaps. While this could be due to our substantial simplification of the magnetic ground state, given the relative weakness of the exchange interactions and associated low ordering points, we reason that the energy gain of the true magnetic ground state would not likely be sufficient to effectively engineer a band gap of several eV. At the same time, one would normally reason that a material with such a large O component and a highly electropositive cation like K would be predisposed towards the highly insulating behavior observed in experiment, so it seems apparent that there is an aspect of the behavior of this material that our theoretical work is not capturing. It remains an open question whether more advanced theoretical treatments, such as dynamical mean-field theory and its extensions, would yield more accurate results here. 

\section{\label{sec:con} Conclusions}

In conclusion, we performed comprehensive bulk and local probe measurements on the physical properties of \KMAO\ to study it's suitability as a candidate Weyl Hydrogen atom. Using salt flux and solid state synthesis techniques, high quality single crystal and powder samples were obtained and shown to order in the reported \Ctc\ structure with no clear evidence of defects or vacancies. Bulk magnetization measurements show two low temperature magnetic transitions at 8 and 4 K. Isothermal measurements and temperature dependent measurements under varying applied fields suggest that the magnetic order is likely ferrimagnetic. Heat capacity measurements reveal only a single lambda like anomaly at $\approx$ 4 K perhaps indicating the 8 K transition is only short-range or partial order. Optically, the obtained crystals were brown and transparent suggesting a relatively large band gap which ellipsometry measurements show to be $\approx$ 4 eV, putting \KMAO\ far away from the desired semi-metallic state. 

Neutron powder diffraction of the low temperature phases identified two ordering vectors \k , with the higher temperature transition exhibiting \ko $=(0,0,0)$ order and the lower temperature transition exhibiting both \ko\ and a second \kt $=(1,0,0)$ thus describing a two-\k\ magnetic order. Magnetic structure solution of the first phase suggests a ferrimagnetic order largely resulting from moments on the lower symmetry Mn2 site. However, the observed peaks are weak and broad consistent with a short-ranged nature. At base temperature, this ferrimagnetic structure from \ko\ superimposes with a second largely AFM structure associated with \kt\ leading to a complex ferrimagnetic structure with a slight modulation of the magnetic moment along the Mn chains and Mn1 moments $\approx$50\%\ larger than those of Mn2.  Both magnetic structures break the $C_{2z}$ symmetry proposed as necessary to stabilize the Weyl nodes predicted for the semi metallic state. However, the \ko\ symmetry analysis allows for a $C_{2z}$ preserving state and so it is perhaps achievable via an external or internal tuning parameter. Revisiting DFT calculations we find expectations for an insulating antiferromagnetic ground state however, with a much smaller gap size than experimentally observed. Our calculations when compared to the experimentally observed ordering, also suggest potential magnetic frustration leading to a suppressed ordering temperature. It is unclear why the material's experimental properties diverge from the predicted band structure indicating more work is needed to understand both the material's properties, including the possibility of local disorder, and accuracy of more advanced theoretical approaches in predicting them. 

Overall, our work shows \KMAO\ as an intriguing material which provides a platform to study 1-D physics, frustrated magnetism, 2-\k\ order, and a test of the ability for first principles calculations to predict the behavior of transition metal compounds. Considering the initial prediction of a topological semi-metallic state it may be possible given the highly flexible alluaudite structure and chemistry, to use several approaches to push \KMAO\ closer to a semi-metallic state with the desired magnetic properties including charge doping, internal pressure, or external pressure encouraging further work on this complex material. 

%

\begin{acknowledgments}
This research used resources of the Advanced Photon Source, a U.S. Department of Energy (DOE) Office of Science User Facility operated for the DOE Office of Science by Argonne National Laboratory under Contract No. DE-AC02-06CH11357. This research used resources at the High Flux Isotope Reactor, a DOE Office of Science User Facility operated by the Oak Ridge National Laboratory. The beamtime was allocated to HB-2A on proposal numbers IPTS-33093 and IPTS-30331. The beamtime was allocated to HB-2C on proposal number IPTS-31172. 
The research is partly supported by the U.S. DOE, BES, Materials Science and Engineering Division. This research used resources of the Compute and Data Environment for Science (CADES) at ORNL, which is supported by the Office of Science of the U.S. DOE under Contract No. DE-AC05-00OR22725. The bulk probe measurements performed in the Materials Science Division at Argonne National Laboratory were supported by the U.S. Department of Energy, Office of Science, Basic Energy Sciences, Materials Science and Engineering Division.
The submitted manuscript has been created by UChicago Argonne, LLC, Operator of Argonne National Laboratory (“Argonne”). Argonne, a U.S. Department of Energy Office of Science laboratory, is operated under Contract No. DE-AC02-06CH11357. The U.S. Government retains for itself, and others acting on its behalf, a paid-up nonexclusive, irrevocable worldwide license in said article to reproduce, prepare derivative works, distribute copies to the public, and perform publicly and display publicly, by or on behalf of the Government. The Department of Energy will provide public access to these results of federally sponsored research in accordance with the DOE Public Access Plan. http://energy.gov/downloads/doe-public-access-plan

\end{acknowledgments}

\nocite{*}

\bibliography{K2Mn3As3O12}

\end{document}